\documentclass[bibyear]{aa}
\usepackage{graphicx}
\usepackage[version=3]{mhchem}
\usepackage{diagbox}
\usepackage{natbib}
\bibpunct{(}{)}{;}{a}{}{,} % to follow the A&A style

\begin{document}

  \title{Simultaneous multicolour optical and near-IR transit photometry of GJ 1214b with SOFIA} 
% * <claudia.dreyer@dlr.de> 2017-03-31T15:05:49.800Z:
% 
% to shorten the title:  "Simultaneous multicolor optical and near-IR transit photometry of GJ 1214b with  SOFIA"
% instruments will be introduced in abstract and text
% 
% 
% ^.
%  \subtitle{}
   \titlerunning{GJ 1214b photometry with SOFIA}

   \author{
D. Angerhausen\inst{1,13,14},
C. Dreyer\inst{2},
B. Placek\inst{3},
Sz. Csizmadia\inst{2},
Ph. Eigm\"uller\inst{2},
M. Godolt\inst{2,11},
D. Kitzmann\inst{1},
M. Mallonn\inst{8},
E. E. Becklin \inst{5,9},
P. Collins\thanks{deceased, January 16, 2017} \inst{4},
E. W. Dunham\inst{4},
J.L. Grenfell\inst{2},
R.T. Hamilton \inst{9},
P. Kabath\inst{6},
S. E. Logsdon\inst{5},
A. Mandell\inst{13},
G. Mandushev \inst{4},
M. McElwain\inst{13},
I. S. McLean\inst{5},
E. Pfueller\inst{7},
H. Rauer\inst{2,11},
M. Savage \inst{12},
S. Shenoy\inst{9},
W. D. Vacca\inst{9},
J. E. Van Cleve \inst{10},
M. Wiedemann\inst{7},
J. Wolf \inst{7}
     }
    \authorrunning{Angerhausen et al.}
    \institute{Center for Space and Habitability, University of Bern, Sidlerstrasse 5, 3012 Bern, Switzerland 
   \email{daniel.angerhausen@csh.unibe.ch}
    \and %2
    Department of Extrasolar Planets and Atmospheres, Institute of Planetary Research, German Aerospace Center, Rutherfordstrasse 2, 12489 Berlin, Germany    %\email{claudia.dreyer@dlr.de}
    \and %3
    Department of Sciences, Wentworth Institute of Technology, Boston, MA 02115, USA
    \and %4
    Lowell Observatory, 1400 West Mars Hill Road, Flagstaff, AZ 86001, USA
    \and %5
    Department of Physics and Astronomy, University of California Los Angeles (UCLA), 465 Portola Plaza, Los Angeles, CA 90095, USA
    \and %6
    Astronomical Institute ASCR, Fri\v{c}ova 298, 25165, Ondr\v{e}jov, Czech Republic
    \and %7
    Deutsches SOFIA Institut, University of Stuttgart Pfaffenwaldring 29, 70569 Stuttgart, Germany
    \and %8
    Leibniz-Institut f\"ur Astrophysik Potsdam (AIP), An der Sternwarte 16, D-14482 Potsdam, Germany
    \and %9
    USRA-SOFIA Science Center, NASA Ames Research Center, Moffett Field, CA 94035, USA
    \and %10
    SETI Institute, 1533 16th Place, Longmont, Colorado 80501, USA
    \and %11
    Centre for Astronomy and Astrophysics, TU Berlin, Hardenbergstrasse 36, 10623 Berlin, Germany
        \and %122
    University of California Observatories, University of California, Santa Cruz, 1156 High Street, Santa Cruz, Ca. 95064
     \and %13
    USRA-NASA postdoctoral fellow, Exoplanets and Stellar Astrophysics Laboratory, Code 667, NASA Goddard Space Flight Center Greenbelt, MD 20771, USA
    \and %14
    Blue Marble Space Institute of Science,
1001 4th ave, Suite 3201
Seattle, Washington 98154
USA}

  \date{review, July 2017}

% \abstract{}{}{}{}{} 
% 5 {} token are mandatory
  \abstract
   	{The benchmark exoplanet GJ 1214b is one of the best studied transiting planets in the transition zone between rocky Earth-sized planets and gas or ice giants. This class of super-Earth/mini-Neptune planets is unknown in our Solar System, yet is one of the most frequently detected classes of exoplanets. Understanding the transition from rocky to gaseous planets is a crucial step in the exploration of extrasolar planetary systems, in particular with regard to the potential habitability of this class of planets.}
    {GJ 1214b has already been studied in detail from various platforms at many different wavelengths. Our airborne observations with SOFIA add information in the Paschen-$\alpha$ cont. $1.9\,\mathrm{\mu m}$ infrared wavelength band, which is not accessible by any other current ground- or space-based instrument due to telluric absorption or limited spectral coverage.}
    {We used FLIPO and FPI+ on SOFIA to comprehensively analyse the  transmission signal of the possible water-world GJ 1214b through photometric observations during transit in three optical and one infrared channels.}
   {We present four simultaneous light curves and corresponding transit depths in three optical and one infrared channel, which we compare to previous observations and state-of-the-art synthetic atmospheric models of GJ 1214b. The final precision in transit depth is between 1.5 and 2.5 times the theoretical photon noise limit,  not sensitive enough to constrain the theoretical models any better than previous observations. This is the  first exoplanet observation with SOFIA that uses its full set of instruments
available to exoplanet spectrophotometry.  Therefore we use these results to evaluate SOFIA’s potential in this field and suggest future improvements.}
 {} 
  
  \keywords{Planets and satellites: individual: GJ 1214b -- Planets and satellites: atmospheres -- Techniques: photometric -- Methods: observational -- Methods: data analysis -- Stars: activity}
	  % A\&A Keywords   http://www.aanda.org/index2.php?option=com_content&task=view&id=170&Itemid=184

  \maketitle

%
%
%________________________________________________________________
\section{Introduction}

\subsection{GJ 1214b}

Since the detection of the transiting Super-Earth GJ 1214b its true nature has been the subject of great interest and is still strongly debated. 
Discovered within the \textit{MEarth} program \citep{nutzman08} by \citet{charbonneau09} GJ 1214b has a radius only 2.7 times larger, while its mass is 6.5 times that of Earth. It transits a nearby ($12.95 \pm 0.9$ pc) M4.5V star with an orbital period of 1.5804 days and has a semi-major axis of 0.0197 AU \citep{harpsoe13}. This results in a planet-to-star flux ratio comparable to that of a Jupiter-sized planet orbiting the Sun, which makes it one of the few super-Earth atmospheres that can currently be investigated with transit spectroscopy. Previous mass and radius measurements of GJ 1214b can be explained by various interior structure and composition models, e.g. with large or small water inventories, depending on the assumptions made for the planetary atmosphere (as e.g. described by \citet{rogers10, nettelmann11}). Differentiating between the various water or hydrogen-dominated atmospheres could help distinguish between these
interior and composition scenarios and would help set constraints on the formation history of this planet, which has no counterpart in the Solar System.
This degeneracy can be broken, and the composition of the planetary atmosphere constrained, by observing transmission spectra of planet's atmosphere. Such measurements have been performed for GJ 1214b by several groups using either space telescopes such as the HST \citep{berta11, kreidberg14} and Spitzer \citep{desert11, fraine13} or ground-based facilities such as VLT \citep{bean10, bean11}, CFHT \citep{croll11}, GTC \citep{murgas12,wilson14}, IRSF \citep{narita13a}, and LBT \citep{nascimbeni15}. First observations suggest a flat transmission spectrum at short wavelengths for GJ 1214b \citep{bean10}, which is consistent with an atmosphere composed of at least 70\% \ce{H_2O} by volume. An alternative interpretation of the data is that GJ 1214b's atmosphere is hydrogen dominated. In this case high-altitude clouds or hazes diminish molecular absorption features at short wavelengths more effectively than at longer wavelengths, whereas a water-rich atmosphere would produce a flat spectrum across all wavelengths. Some measurements support the featureless spectrum (\citet{crossfield11}, Keck-NIRSPEC; \citet{desert11}, Spitzer-IRAC) whereas other observations indicate large features around the g-band and the K-band which would imply a \ce{H_2}-rich atmosphere (\citet{croll11}, CFHT-WIRCam; \citet{demooij12}, INT-WFC/ESO-GROND/NOTCam/WHT-LIRIS; \citet{demooij13}, VLT-FORS/WHT-ACAM/INT-WFC; \citet{teske13}, Kuiper 1.55 m telescope/STELLA-WiFSIP).\\
The emerging class of super-Earths/mini-Neptunes are likely to be common in the Galaxy \citep[e.g.][]{marcy14}. Furthermore, these planets  represent an important stepping-stone in the data-driven pathway towards characterising Earth-like exoplanets.
GJ 1214b has been the subject of many atmosphere modeling studies, focusing on e.g. the impacts of clouds or chemistry on the spectral appearance and characterisation of the planet \citep{miller-ricci10, miller-ricci_kempton12, howe12, menou12, benneke13, morley13, hu14}. Detailed studies considering the formation of clouds and atmospheric dynamics that aim to investigate the formation and nature of clouds and hazes in the atmosphere of GJ 1214b include, for example, \cite{morley15} or \cite{charnay15}.
  
  In order to distinguish between a water-dominated atmosphere (larger mean molecular weight/smaller scale height) and a hydrogen-dominated atmosphere (smaller mean molecular weight/larger scale height), we performed a SOFIA primary transit observation of GJ 1214b. This transit observation specifically targeted the water band around $1.85\,\mathrm{\mu m}$ using FLITECAM's extremely narrow-band \textit{Paschen-$\alpha$\,cont.} filter centered at $1.90\,\mathrm{\mu m}$ (designed to target the Paschen-$\alpha$ continuum). This wavelength is especially interesting since different results have been observed in the K-band around $2.2\,\mathrm{\mu m}$ and our 'Paschen-$\alpha$ cont.' data point adds another important adjacent data point close to the K-band. %Observations at this wavelength have not yet been carried out since it is not observable from the ground (the Earth atmosphere strongly absorbs at that wavelength right between H and K Band) and no present space mission covers these wavelengths.
 % (HST WFC long edge is at 1.7 . Furthermore, at that wavelength the water dominated \& the hydrogen dominated model atmospheres show strong differences (Fig.\,\ref{fig:gj1214b_spec_lit}) which should be observable.

%                                					 one column figure
%----------------------------------------------------------- 
  %  \begin{figure}
    %   \centering
       %\includegraphics[width=0.48\textwidth]{PICTURES/fpi_gj1214_fov_named.eps}
     %   \caption{xxx}
     %  \label{fig:gj1214b_spec_lit}
  %  \end{figure}
\subsection{Observing exoplanets with SOFIA}
 The measurement close to the  $1.85\,\mathrm{\mu m}$ water band is only possible with SOFIA: telluric absorption almost completely shuts down this band between H and K from ground-based observatories and available space-based telescopes do not cover that wavelength regime. Furthermore, SOFIA provides the only platform for simultaneous optical and infrared observations that are inaccessible from the ground. 
 When conducted from ground-based platforms, spectrophotometric exoplanet observations are significantly affected by the perturbing variations of trace gases, in particular \ce{H_2O}, in the Earth's atmosphere. It was theorised that the airborne platform SOFIA had some unique advantages for this kind of exoplanet research \citep{dunham07, Gehrz10, angerhausen10, angerhausen14, cowan15}. 
 SOFIA's cycle 1 observation of HD~189733b, then demonstrated that SOFIA can overcome the hurdle of changing atmospheric absorption in the optical in absolute photometry without the use of field  stars; \citet{angerhausen15} demonstrated a precision of $\sim$150 ppm in absolute optical photometry of HD~189733b. SOFIA can leverage bright host stars to the fullest and is therefore not limited in S/N by much fainter comparison stars that need to be used from the ground.\\

In the following sections we briefly introduce the instruments on SOFIA that can be used for exoplanet spectrophotometry.

\subsubsection{HIPO}
 The \textit{High Speed Imaging Photometer for Occultations} (HIPO) is a Special Purpose Principal Investigator class Science Instrument (SSI, \citet{dunham04, dunham14}).  HIPO is designed to provide simultaneous high-speed time resolved imaging photometry at two optical wavelengths. 
 %The primary HIPO detectors are e2v CCD47-20 $1024\times 1024$ pixel frame transfer CCDs with plate scales of $0.33''\times 0.33''$ pixels at low resolution and $0.05''\times 0.05''$ pixels at high resolution. 
 The HIPO field of view (FoV) is a 5.6' square, the 8' diagonal of which corresponds to the 8' diameter SOFIA field of view. The filter set includes the Johnson (UBVRI) and Sloan (u'g'r'i'z') filters as well as a filter for methane at 890 nm.
 %A number of readout modes are available allowing the observer to optimise the subframe size, speed, noise, full well capacity and linearity trade off for any particular event.

\subsubsection{FLITECAM}
 The  \textit{First Light Infrared TEst CAMera} (FLITECAM) is a near-infrared imager and grism spectrograph covering the $\sim 1 - 5\,\mu$m range \citep{mclean06, logsdon14}. 
 %FLITECAM's $1024\times 1024$ InSb ALADDIN III array covers an $\sim 8\arcmin$ diameter field of view with a plate scale of 0.475$\arcsec$ pixel$^{-1}$. 
 The full set of available FLITECAM filter pass-bands are listed online in the FLITECAM chapter of the SOFIA Observer’s Handbook. FLITECAM was co-mounted with the HIPO instrument during these observations, a configuration that precluded observations at wavelengths longer than $\sim 4~\mu$m, and reduced the sensitivity at wavelengths longer than $\sim 2~\mu$m, due to high background levels resulting from the warm dichroic and transfer optics.  

\subsubsection{FPI+}
 On SOFIA the light passes through the telescope's dichroic tertiary mirror (25\% and 45\% reflectivity for the B and z' bandpasses) to the \textit{Focal Plane Imager }(FPI+, \citet{pfueller16}). 
 %Most of the visual light passes the tertiary beam splitter before it is reflected into the Nasmyth tube by the fully-reflective tertiary. The peak transmission of the tertiary beam splitter is at 570 nm with 64\% of the light passing the mirror. A significant amount of visual light is not transmitted but rather either absorbed or reflected along with the longer, infrared wavelengths. However, in the range between 480 nm to 800 nm where the visual-light CCD cameras are most sensitive, more than 50\% of the light is transmitted. 
 The FPI+ contains a highly sensitive and fast EM-CCD camera. Its images are primarily used for tracking but can also be stored without disrupting the tracking process and in parallel with measurements of the instruments mounted to the telescope. With the released call for proposals for the SOFIA observing cycle 4 (2015), the FPI+ was made available for proposals as a facility science instrument for observations in 2016 and thereafter.

%
%
%__________________________________________________________________
\section{Observation}
 This joint US-German Cycle 2 GI program (US-proposal: \citet{angerhausen13}; German-proposal: \citet{dreyer13}) was performed on SOFIA’s flight number 149 on UT February 27, 2014.

\begin{figure}[ht]
       \centering
       \includegraphics[width=0.48\textwidth]{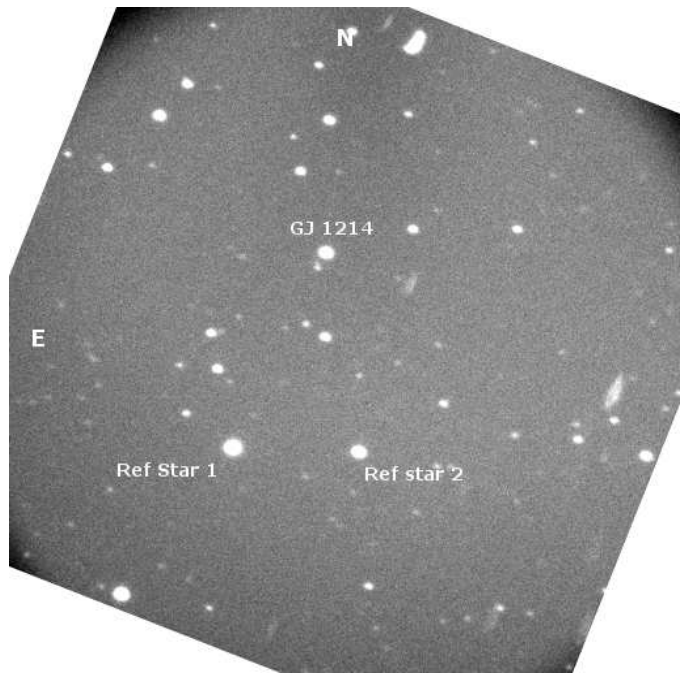}
       \caption{SOFIAs $8'\times 8'$ FPI+ field of view showing our target GJ 1214 and reference stars}
       \label{fig:skymap}
    \end{figure}

 We observed the transit of the exoplanet GJ 1214b using the photometry mode of FLITECAM and HIPO in the `FLIPO' configuration in order to perform differential aperture photometry of the target and a bright comparison star (see Table \ref{tab:comp_stars} and Figure \ref{fig:skymap}). GJ 1214b was monitored during one 52 min transit plus ca. 70 min before and 10 min after transit for a total of 150 min (including some additional time for setups and calibrations). Observations were simultaneously conducted in two optical HIPO channels: open blue at $0.3 - 0.6\,\mathrm{\mu m}$ and Sloan z' at $0.9\,\mathrm{\mu m}$ and one infrared FLITECAM filter: Paschen-$\alpha$ cont. at $1.9\,\mathrm{\mu m}$. The individual exposure time for all HIPO and FLITECAM frames was 25 sec. 
 Complementary data were also obtained with the optical focal plane guiding camera FPI+ in the Sloan i' band ($0,8\,\mathrm{\mu m}$) with mostly 2.5 sec (but some with 3 sec) exposure time, as it was used for both tracking and data acquisition purposes. The change from 3 sec to 2.5 sec integrations was triggered by an increasing photon count at higher elevations. We chose to take shorter exposures to avoid saturation. We used an open filter for the HIPO blue side in order to cope with the faintness of GJ 1214 at blue wavelengths. This bandpass is defined by the transmission of the atmosphere, telescope, HIPO blue side optics, and the reflection curve of the internal HIPO dichroic reflector with a transition wavelength of $675\,\mathrm{nm}$. The HIPO red filter was selected to avoid possible telluric ozone variability, while the FLITECAM filter was chosen due to its wavelength coverage of a prominent \ce{H_2O} spectral feature that cannot be sampled from ground-based observatories. The filter selected for the FPI+ was an intermediate wavelength between the HIPO channels that is somewhat ozone sensitive. This provided the potential for detecting and removing residual telluric ozone-related systematics from the HIPO z' filter.
  For this observation the FPI+ acquired images with a Sloan i' filter to complement the HIPO blue channel and Sloan z' filter. The Sloan i' filter has a central wavelength of 760 nm with a pass band between 694 nm and 843 nm. The average throughput with this filter is 24.4\% taking into account a simulated atmosphere at flight altitude, the reflectivity and transmittance of all optical elements in the light path through the telescope and the CCD sensor quantum efficiency. In addition to the acquisition of science data, the FPI+ was simultaneously used as a tracking camera to keep the telescope precisely pointed at the target. The tracking accuracy, as measured with the FPI+, was 0.17 arcseconds rms. 
Full frame images ($1024\times 1024$ pixel) were taken with a $2\times 2$ pixel binning which resulted in a spatial resolution of 1.03 arcseconds per pixel and a square field of view of 8.8 arcminutes. The image integration time was set to 2.5 seconds to achieve maximum pixel values at about 65\% full well capacity of the sensor.
The HIPO instrument was operated in \textit{Basic Occultation mode} with full-frame read-out to maximize our field standard possibilities.
 We used simple stare mode to minimize the contribution of systematic errors. In addition, slight defocussing was applied in order to minimize the potential instability problems with telescope guiding and to increase the S/N ratio. Bias and all other frames were taken at $512\times512$. \textbf{Table \ref{tab:observation} gives an overview of the final data set.} 
 
 Due to flight planning constraints, the end of the transit occurred in morning twilight and in the last $\sim$15 minutes the sky brightness gradually increased to about 3.5 times its night-time values. 
 
 In Section \ref{sec:PCA} we describe how we correct for the changing observational parameters in general via a principal component analysis method, in Section \ref{sec:HipoCorr} we describe how we had to correct for the twilight contribution to the HIPO blue channel.
%                                					 one column figure
%----------------------------------------------------------- 
%
%__________________________________________________ one column table
\begin{table}[ht!]
\centering
     \caption[]{GJ1214 and reference stars in SOFIAs FPI+ field of view}
     \label{tab:comp_stars}
      \begin{tabular}{l l r@{.}l c}
      \hline\hline
      \noalign{\smallskip}
      Star & 2MASS &  \multicolumn{2}{c}{K}     &   dist\\
           &       &  \multicolumn{2}{c}{mag}   & (arcmin) \\
      \noalign{\smallskip}
      \hline\hline
      \noalign{\smallskip}
		GJ 1214     & J17151894+0457496 &  8&782  &0\\
		Ref. star 1 & J17152424+0455041 &  8&831  &3.05\\
		Ref. star 2 & J17151760+0455021 & 10&318  &2.81\\
	    \noalign{\smallskip}
      \hline
      \end{tabular}
\end{table}

\begin{table*}[ht!]
\centering
     \caption[]{Observation Summary }
     \label{tab:observation}
      \begin{tabular}{l c c c c}
      \hline\hline
      \noalign{\smallskip}
           & \multicolumn{2}{c}{HIPO} &  FPI+     & FLITECAM    \\
           \noalign{\smallskip}   
           \cline{2-3}
          \noalign{\smallskip}     
           & HIPO-blue & HIPO-red   \\
           \noalign{\smallskip}
      \hline\hline
      \noalign{\smallskip}
		observation time [UTC] & \multicolumn{4}{c}{2014-02-27, 10:00:55 -- 13:43:33}\\
        \noalign{\smallskip}
        \hline
        \noalign{\smallskip}
        pass-band$^*$                     & open blue & SDSS z' &  SDSS i'  & Pa $\alpha$ cont.\\
        $\lambda_\mathrm{eff}\,[\mu m]^*$ & 0.3-0.7   & 0.89    & 0.76      & 1.90\\
        band-width $[\mu m]^*$            & 0.4       & 0.23    & 0.15      & 0.02\\
        exposure time [sec]	              &  25       &  25     & 2.5/3     &  25\\
        \noalign{\smallskip}
        \hline
        \noalign{\smallskip}
        No. of frames	    & 459     & 448  & 4786 (2882/1904) & 433\\
        image [px] & \multicolumn{2}{c}{512 x 512} &  512 x  512 & 1024 x 1024\\
        dark [px]  & \multicolumn{2}{c}{512 x 512} &  512 x  512 &  --\\
        flat [px]  & \multicolumn{2}{c}{512 x 512} &  512 x  512 &  --\\
		bias [px]  & \multicolumn{2}{c}{512 x 512} & 1024 x 1024 &  --\\
        \noalign{\smallskip}
      \hline
      \end{tabular}
         \begin{list}{}{}
          \item \textbf{Notes.} $ ^*$taken from Sofia Observer's Handbook for Cycle 2: v2.1.2
     \end{list}
\end{table*}

%
%
%__________________________________________________________________
\section{Data reduction and analysis}\label{sec:datareduction}

\subsection{Light curve extraction}
 Standard data reduction was applied to the data taken with HIPO (red and blue) and FPI+. This includes bias and dark subtraction, and in the case of FPI+ also flat field correction. For FLITECAM we did not acquire bias frames, as bias contributions are generally very low for this type of NIR detector array. Furthermore it is complicated and time consuming to obtain a reliable flat field on such a narrow band filter as the 1.9 $\mu m$ Paschen-$\alpha$ continuum filter. Since it was not possible to take long enough exposures during this campaign, we used K-band flat fields taken on the same flight before our observation run, which, however, did not improve the photometric precision significantly. Similarly, dark subtraction did not show any improvement in photometric precision. Additionally we corrected for the sky background using dithered images taken during the observation run. Stars were detected with the Source Extractor by \citet{bertin96}. Aperture photometry was applied using IRAF \citep{tody93} / DAOPHOT \citep{stetson87} using circular apertures. The optimal aperture radius with the lowest noise level was found to be 6 pixels  for FLITECAM and 12 pixel for HIPO and FPI+. As part of the DAOPHOT routine, an annulus around the target was used to estimate the sky background in each exposure. Next to GJ 1214, we extracted the light curves of two additional bright stars within our field of view (see Table \ref{tab:comp_stars} and Figure \ref{fig:skymap}). To identify stars in the images we calculated a rough astrometric solution for each image using data provided by Astrometry.net \citep{barron08}. In the FLIPO setup SOFIA does not provide an image rotator to compensate field rotation during long integrations. This introduces a rotation of the images over time. Due to SOFIA's unique setup, the telescope must periodically undergo so-called `Line-of-Sight (LOS) rewinds'. The required frequency of LOS rewinds depends on rate of field rotation experienced by the target, which is a complex function of the position of the target in the sky relative to that of the aircraft heading. These need to be carefully timed with regard to the transit observation, to not interfere with, e.g., ingress or egress. While we kept the target star, GJ1214, in boresight, the comparison stars moved over the CCD due to this field rotation. This is one of the main factors introducing systematic noise and limiting the photometric precision of the instrument and is another reason why reliable flat fields are crucial for this kind of time series observation. After comparing the light curves, the brighter star was selected as comparison star to correct for first order systematic effects present in all light curves.

\subsection{Observational parameters in the airborne environment} \label{sec:obsparam}
Photometric observations from an airborne platform like SOFIA differ from ground-based observations. While ground-based photometry suffers from systematic errors induced by e.g. air mass or local weather changes, photometric observations with SOFIA also correlate with changes in flight parameters such as Mach-number or air density. 

\begin{figure*}
      \centering
      \includegraphics[width=0.95\textwidth]{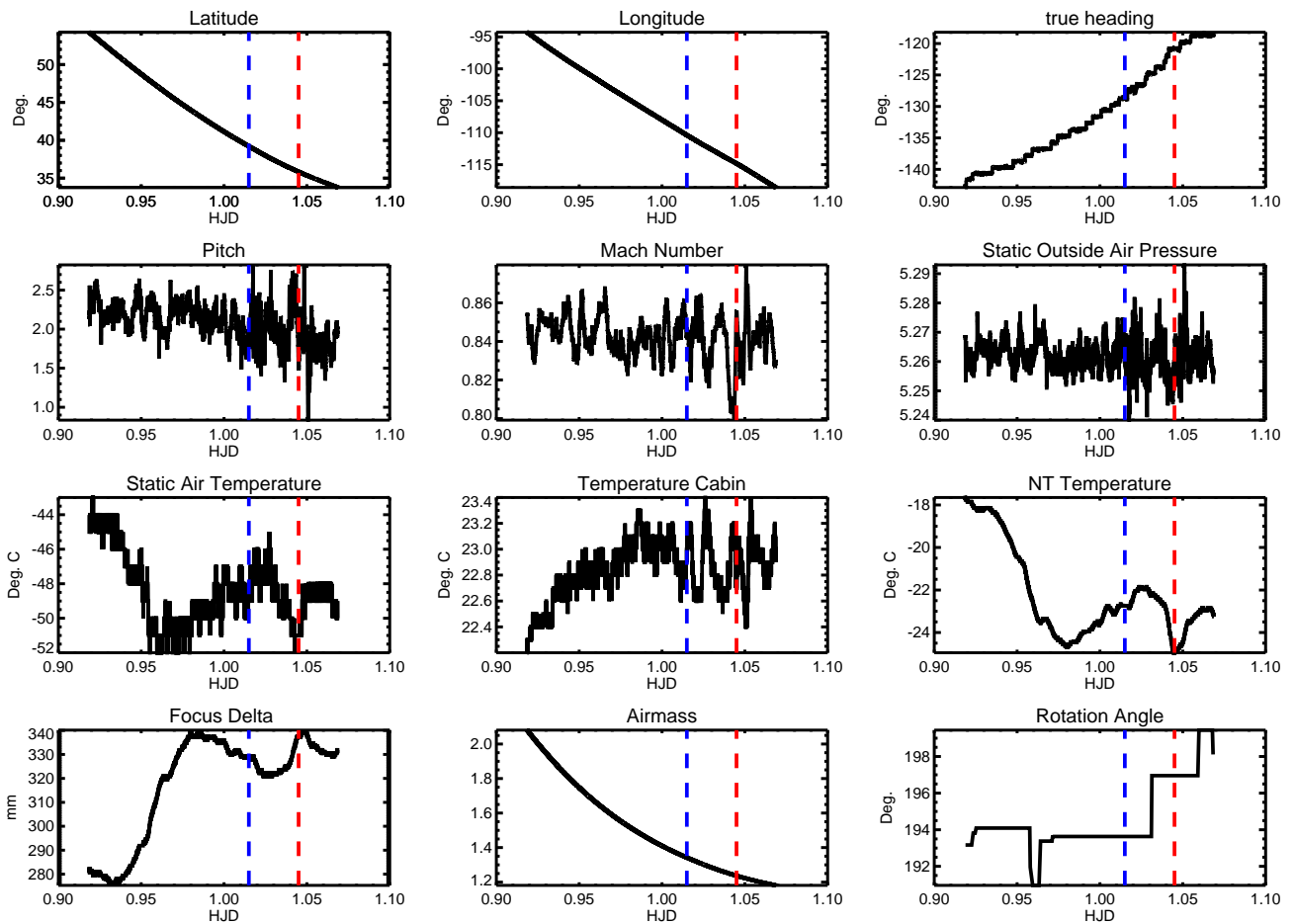}
      \caption{Sample time series of some observational parameters. While parameters like the plane heading or pitch  should not influence the photometry, they illustrate the mobile airborne environment. The beginning of ingress and end of egress are marked in blue and red vertical lines. }
      \label{fig:hk_plot}
    \end{figure*}

We used the housekeeping data taken during our observation to parameterize the time dependence of our observational environment. Figure \ref{fig:hk_plot} shows time series of some selected observational parameters, some of them unique to the airborne environment. Many of these parameters are mutually correlated. In order to overcome these degeneracies we performed a principal component analysis on all available parameters to produce a set of linearly independent time series to eventually decorrelate the raw light curves (see Section \ref{sec:PCA}).
 
%                                 					Two column figure
%----------------------------------------------------------- 

%
%
%______________________________________________________________________
\subsection{Light-curve and noise modeling} \label{sec:modelling}

In this section we present two methods that were used to fit the resulting light curves. In the second case we also present a method to correct for the aforementioned systematics induced by the airborne environment.

 \subsubsection{TLCM applied to the raw lightcurves} \label{sec:TLCM_modelling}
 The main features of the `Transit Light Curve Modeller (TCLM)' code are described in \citet{csizmadia11, csizmadia15}. Therefore we repeat only the most important  pieces of information about TLCM here. The TLCM uses the formalism of \citet{mandel02}, which is based on spherical star and planet shapes, to fit the light curves. To optimize the fit, first a genetic algorithm-based Harmony Search \citep{geem01} was performed, then Amoeba refined the fit \citep{press92}. Finally we used Simulated Annealing (SA) (\citet{press92} and references therein) for error  estimation as well as to better monitor the possible parameter correlations. The SA chain consisted of $10^5$ steps. The SA process is quite similar to the Markov Chain Monte Carlo (MCMC). However, in SA the control parameter (the so-called `temperature' of the  Metropolis-Hastings procedure of MCMC) is continuously and slowly decreased. When this `temperature' is very small, then SA will be similar to a simple random walk, and when it is large then it is equivalent to MCMC. We decreased this temperature by 1\% after every 2000 steps starting from such a value that in this way we reached that the overall acceptance rate was around 30\%.

 The host star, GJ1214, is a chromospherically active M-dwarf \citep{nascimbeni15}. Thus spot activity may affect the light curve fit in several different ways. The following effects seem to be important:
 \begin{itemize}
  \item[a)] The rotational modulation of the stars caused by spots and stellar rotation yields a long-term oscillation of the light curve that has a much longer time-scale (days/weeks) than the length of the transit or our observational window. This effect was removed with a parabolic baseline fit.
  \item[b)] Spot-crossing during the transit \citep[see e.g.][]{sanchis-ojeda11, silva-valio11} severely affected other observations of GJ 1214b’s transit \citep[e.g.][]{bean11}. However, there is no clear sign of spot-crossing at our epoch of observation or it is lost in the noise.
  \item[c)]Spots that do not cross and/or are polar spots will cause changes in the observable limb darkening coefficients \citep{csizmadia13b}. In addition, the theoretically predicted limb darkening coefficients, especially the recent tables, have not been observationally verified. That is why, following the recommendation of \citet{csizmadia13b} and \citet{espinoza15}, we adjusted the limb darkening coefficients.
 \end{itemize}

 The long-term behavior of the stellar variability and the estimated effect of the stellar spots on the systematic and random  errors in the derived planet-to-stellar radius ratios which were not removed by the baseline-fit are discussed separately  in Section \ref{sec:activity}.

 The TLCM-based light curve modeling was carried out by fixing the scaled semi-major axis at $a/R_s = 14.97$ and the impact parameter at b=0.277026. These values were chosen to match the values used by other investigators \citep[e.g.][]{caceres14, desert11, croll11, bean11, demooij12, murgas12, narita13a, narita13b, fraine13, teske13} so that our results are more readily comparable to those works. 
 
 We also note that the eccentricity of GJ 1214b is not well constrained (e.g.~\citet{charbonneau09} gives only an upper limit for eccentricity of $e<0.27$). The impact of eccentricity on the light curve fit is not investigated by other authors, who all assume a circular orbit. However, eccentricity has an effect on the speed of the planet during transit and thus on $a/R_s$. Therefore the stellar density measured from the transit duration is somewhat approximate. Consequently, the stellar parameters should also be considered approximate until the eccentricity has been established. Since we have only one photometric transit measurement and no additional radial velocity follow-up data, we are also not in the position to further constrain the eccentricity. Therefore, we decided to use a circular orbit for the fit – as other authors do – because the analysis can be repeated later if a significant eccentricity is found. We urge the community to collect more radial velocity data points to finally close the eccentricity issue of GJ 1214b.

 Notice that the \citet{mandel02} formalism calculates the planet-star mutual distance projected to the sky as
 
 \begin{equation}
   \delta = a / R_s (\cos \Omega t + sin\Omega t \sin i),~~~ \Omega = 2\pi \frac{t-E}{P}
 \end{equation}
 
 where $P$ is the orbital period, $E$ is the epoch, $t$ is the time and $i$ is the inclination. The equation above is valid for circular orbits. 
 This can be easily generalized to eccentric orbits (e.g. \cite{russell12, gimenez06}):
 
 \begin{equation}
   \delta = a/R_s \frac{1-e^2}{1+e \cos v} \sqrt{1 - \sin^2 i \sin(v+\omega)}
 \end{equation}
 
 where $e$ is the eccentricity, $\omega$ is the argument of periastron, $v$ is the true anomaly calculated from the solution of the  Kepler-equation.

 However, as we mentioned, the eccentricity is not well-known for GJ 1214b and this may significantly affect the end-result. Therefore we used the following equation instead of the one recommended by \cite{mandel02} to describe the sky-projected star-planet distance (\cite{csizmadia13a}):
 \begin{equation}
   \delta = a/R_s \times \sqrt{b^2 + (t-E) \cdot ((1+k)^2 - b^2) / P}
 \end{equation}
The \citet{csizmadia13a} equation is based on the assumption that the planet moves with constant projected velocity during transit. We then interpolate the planet's motion linearly.This assumption is quite good even for high eccentricities and close-in orbits where the the transit light curve would become asymmetric due to the slowly changing projected velocity. However, at the present level of photometric accuracy the asymmetry should not be taken into account \citep[c.f.][]{moutou09} who did not find this asymmetry in the transit of HD 80606b whose eccentricity is 0.93).

 Our free parameters were: four planet-to-stellar radius ratios (one for each of the four passbands we observed in), the four corresponding u+ and u- limb darkening coefficient combinations, and the four epochs of observations (again, one for each passband). The limb darkening combinations were defined as $u+ = u_1 + u_2$ and $u- = u_1 - u_2$ where $u_1$ is the linear and $u_2$ is the quadratic term of the quadratic limb darkening law. According to \citet{brown01} and \citet{pal08}, such combinations are less sensitive to degeneracies between the coefficients. We decided to leave these parameter combinations as free parameters, because no theoretical limb darkening calculations are available for the instruments and passbands we used. In addition, the host star is a convective, active M-dwarf and theoretical  calculations do not include the stellar spots so far nor the probable exciting granulation pattern of small stars. We divided the data by a parabola whose coefficients were fitted simultaneously with the light curve parameters. This parabola served as our baseline-corrections to remove any stellar activity signal or long time-scale instrumental/air mass effects. 

 A simultaneous fit to the different data sets would have the advantage that the number of free parameters are decreased, because the wavelength-independent parameters would be the same for every data set. However, the epoch of observation is not necessarily the same in all colours because of the distribution of data points or because the planetary atmosphere is asymmetric. A planet with non-spherical atmosphere may have different atmospheric density and thus atmospheric transparency causing slightly asymmetric transit shape (e.g. if it loses its atmosphere). If the number of data points are not symmetric to the mid-point of the transit, e.g. there is an unfortunately placed gap, then the timing error increases and the fitted mid-point can be shifted (e.g. Csizmadia et al. 2010). Therefore we fitted the four epochs but we found they are in good agreement with each other (Table 2). In total, we had 28 free parameters for the four pass bands: four epochs, four planet-to-stellar radius ratios, two times four limb darkening coefficients combinations, and four times three coefficients of the parabola.
 
\begin{figure}[ht] 
\centering
\includegraphics[width=0.48\textwidth]{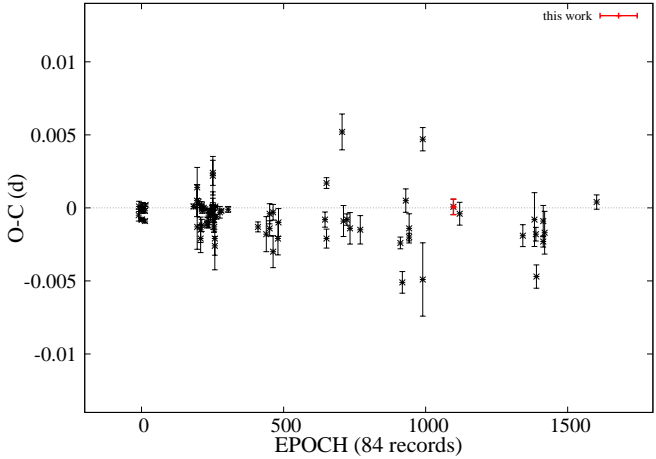} 
\caption{O-C diagram of GJ 1214b. The red point marks our measurement. Black points are taken from \citep{2010NewA...15..297P} using all (professional and citizen science) data from the Exoplanet Transit Database. For example the point at O-C=0.005 comes from one of the TRESCA light curves.  Our data is consistent with other measurements finding no significant long-term variations in transit timing.}\label{fig:TLCM_O-C}
\end{figure}

Using the following ephemeris\footnote{http://var2.astro.cz/ETD}:
 \begin{equation}
 \mathrm{Transit_N} = \mathrm{HJD} 2454980.748795 + 1.58040482 \times N
 \end{equation}
and transforming the observed transit times of Table \ref{tab:lc_fits} into {\it HJD}s, we found that we observed transit N=1098 and the corresponding $O-C$ value is $+0.00007\pm0.00053$ days (cf.\,Fig.\,\ref{fig:TLCM_O-C}), so it is in perfect agreement with the ephemeris.

\begin{figure*}[ht!]
 \centering
  \includegraphics[width=0.48\textwidth]{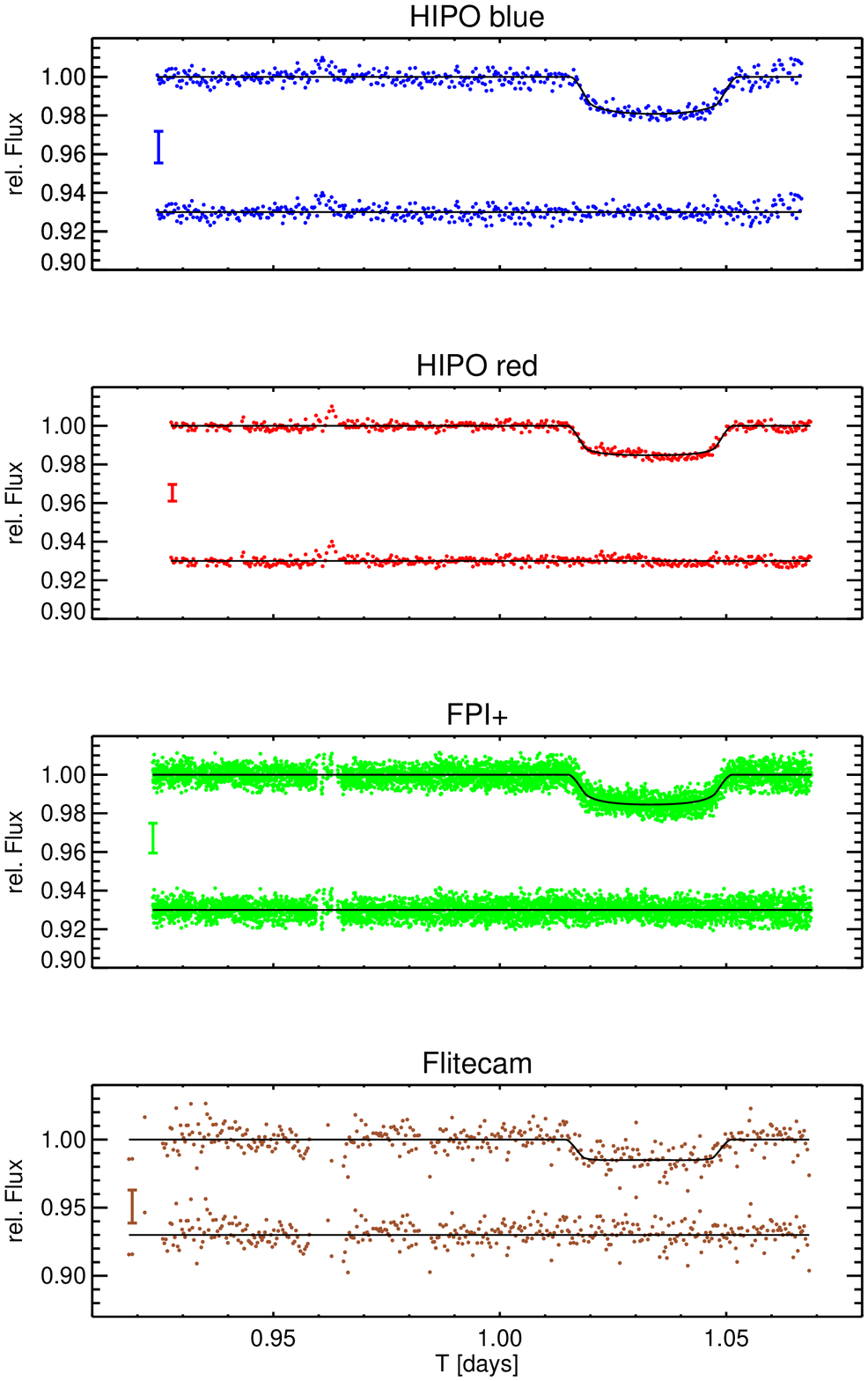}  
     \includegraphics[width=0.48\textwidth]{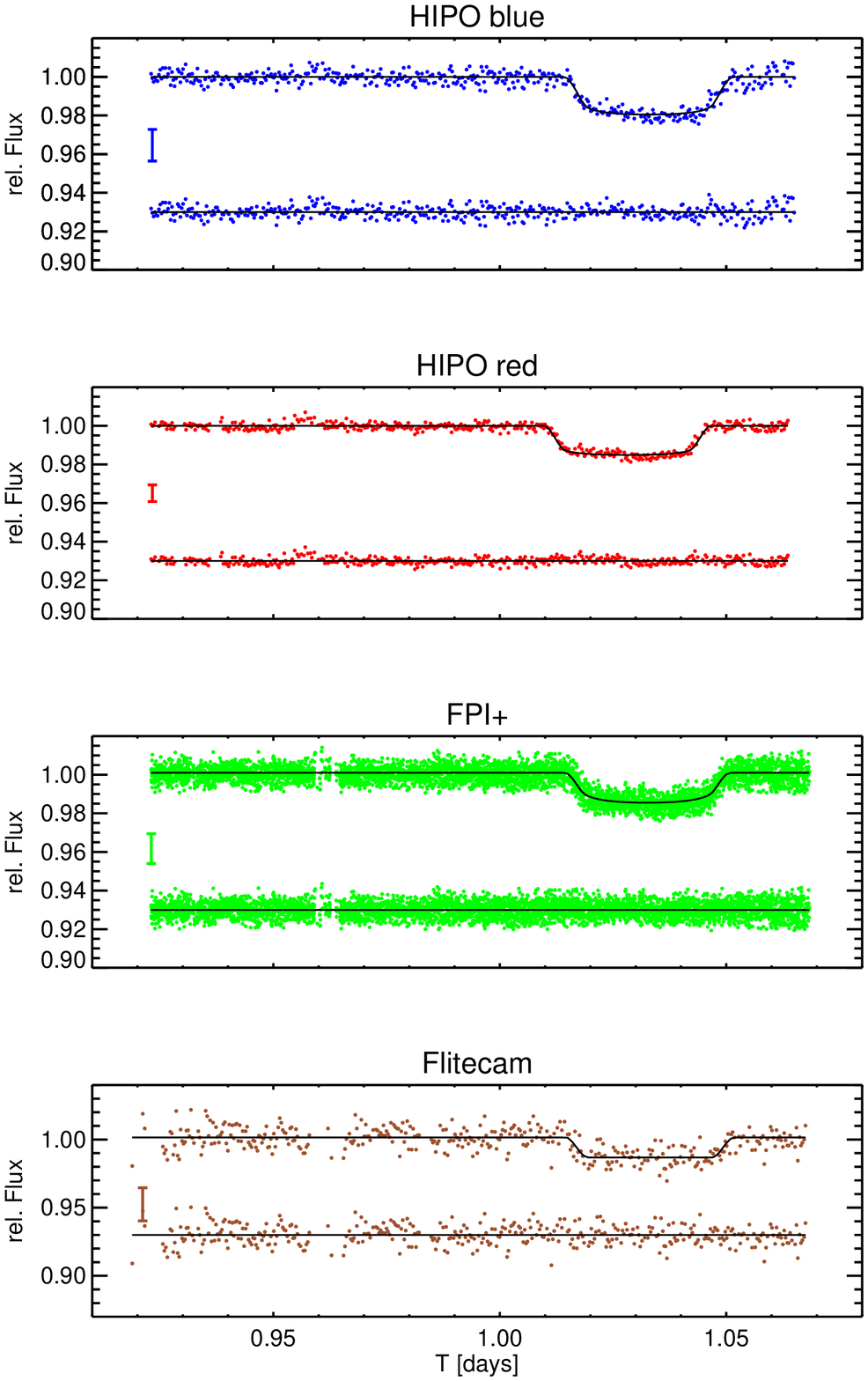}  
   \caption{Left: Results of the TLCM fit to the raw data including a polynomial first order noise model. Right: Final EXONEST fits to the data reduced using a principal component analysis noise model.}
   \label{fig:lc_fits}
\end{figure*}

 The results of the fit are shown in tabular form in Table \ref{tab:lc_fits} and are visualized in Figure \ref{fig:lc_fits}.

\begin{table*}
    \caption[]{Results of the light curve modeling. The epochs are relative to JD 2456715.0. }
     \label{tab:lc_fits}
      \begin{tabular}{l  r@{.}l@{$\ \pm\ $}l  r@{.}l@{$\ \pm\ $}l  r@{.}l@{$\ \pm\ $}l   r@{.}l@{$\ \pm\ $}l r@{.}l@{$\ \pm\ $}l}
      \hline\hline
     \noalign{\smallskip}   
                     &\multicolumn{9}{c}{HIPO} & \multicolumn{3}{c}{FPI+}  & \multicolumn{3}{c}{FLITECAM}\\
     \noalign{\smallskip}   
     \cline{2-11}
     \noalign{\smallskip}   
			         & \multicolumn{3}{c}{HIPO-blue}	
			         & \multicolumn{3}{c}{HIPO-blue, short}
	                 & \multicolumn{3}{c}{HIPO-red}\\
%			         & \multicolumn{6}{c}{open blue = (0.3-0.6) $\mu m$}	
%			         & \multicolumn{3}{c}{Sloan z'= 0.9 $\mu m$} 
%			         & \multicolumn{3}{c}{Sloan i'= 0.8 $\mu m$} 
%			         & \multicolumn{3}{c}{Pa $\alpha$ = 1.9 $\mu m$}\\
      \noalign{\smallskip}
      \hline
      \noalign{\smallskip}
			\multicolumn{16}{l}{Photon noise limited} \\
			sensitivity [ppm/min] & \multicolumn{6}{c}{1000} & \multicolumn{3}{c}{400} & \multicolumn{3}{c}{500} &\multicolumn{3}{c}{1000}\\
			\noalign{\smallskip}
			ppm per exposure      & \multicolumn{6}{c}{1600 (25 sec)} & \multicolumn{3}{c}{650 (25 sec)} & \multicolumn{3}{c}{2500/2220 (2.5/3 sec)} &\multicolumn{3}{c}{1600 (25 sec)}\\
			\noalign{\smallskip}

      \hline
      \noalign{\smallskip}
		P [days]	 & \multicolumn{15}{c}{$1.5804055929$ (fixed)} \\
		$a/R_s$      & \multicolumn{15}{c}{$14.9749$ (fixed)} \\
     	i [degree]   & \multicolumn{15}{c}{$88.94$ (fixed)} \\
	  	e            & \multicolumn{15}{c}{$0$     (fixed)} \\
		b            & \multicolumn{15}{c}{$0.27702737$ (fixed)} \\
	  \noalign{\smallskip}
      \hline\hline
	  \noalign{\smallskip}
      \multicolumn{15}{l}{\textbf{TLCM fitting with polynomial correction}}\\
	  \noalign{\smallskip}
      \hline
      \noalign{\smallskip}
	    $k=R_p/R_s$	   & 0&1281  & 0.003  & 0&1184   & 0.0189       & 0&1156   & 0.0023 & 0&1133  & 0.0029 & 0&1203  & 0.0046\\
	    u+	    	   & 0&541   & 0.293  & \multicolumn{3}{c}{$ $} &  0&572   & 0.19   & 0&747   & 0.293  & 0&35    & 0.30\\
	    u-         	   & 0&497   & 0.81   & \multicolumn{3}{c}{$ $} & -0&189   & 0.35   & 0&143   & 0.73   & -0&71    & 0.83\\
		epoch          & 1&03419 & 0.0005 & \multicolumn{3}{c}{$ $} &  1&03304 & 0.0003 & 1&03313 & 0.0003 & 1&0328 & 0.0005\\
	  \noalign{\smallskip}
      \hline
      \noalign{\smallskip}
			$\chi^2$ of the fit &  \multicolumn{15}{c}{$ 1.2040 $ }\\   
      \noalign{\smallskip}
      \hline\hline			\noalign{\smallskip}
			\multicolumn{15}{l}{\textbf{EXONEST fitting with principal component noise correction$^*$}}\\
			\noalign{\smallskip}
      \hline
			\noalign{\smallskip}
		$k=R_p/R_s$    & 0&1288 & 0.0028 & 0&1225 & 0.0017          & 0&1156 & 0.0026 & 0&1107 & 0.0011 & 0&1215&0.005\\
	    $u_1$		   & 0&704  & 0.292  & \multicolumn{3}{c}{$ $}  & 0&677  & 0.294  & 0&864  & 0.228  & 0&253&0.186\\
	    $u_2$          & 0&011  & 0.280  & \multicolumn{3}{c}{$ $}  &-0&170  & 0.280  &-0&481  & 0.220  & 0&365&0.254\\
     		\noalign{\smallskip}
      \hline
      \noalign{\smallskip}
			$\chi^2$ of the fit &  \multicolumn{15}{c}{$ 1.3245 $} \\   
      \noalign{\smallskip}
%       \hline\hline	
% 			\noalign{\smallskip}
% 			\multicolumn{15}{l}{\textbf{EXONEST fitting with Gaussian Process noise correction}}\\
% 			\noalign{\smallskip}
%       \hline
% 			\noalign{\smallskip}
% 		$k=R_p/R_s$		 &  0&128862&0.002413  &  0&122537& 0.001651 &  0&115579&0.001885  &   0&110875&0.001009  &   0&109878&0.004804\\
% 	    $u_1$			 &  0&709386&0.1244620 &   &      &          &  0&619210&0.1132590 &   0&907277&0.047650  &   0&1254030&0.097847\\
% 	    $u_2$       	 & -0&354763&0.4193730 &   &      &          & -0&474784&0.355362  &  -0&737443&0.188775  &   0&505217&0.389610\\
% 			\noalign{\smallskip}
%       \hline
%       \noalign{\smallskip}
% 			$\chi^2$ of the fit &  \multicolumn{15}{c}{$ 1.3517 $ } \\   			
%     	\noalign{\smallskip}
%       \hline\hline	
% 			\noalign{\smallskip}
% 			\multicolumn{15}{l}{\color{red}{Average -- median of the three fits}}\\
% 			\noalign{\smallskip}
%       \hline
% 			\noalign{\smallskip}
% 		$k=R_p/R_s$	 & & & & & & & & & & & & & & &\\
% 	    %$u_1$		 & & & & & & & & & & & & & & &\\
% 	    %$u_2$     	 & & & & & & & & & & & & & & &\\
% 			\noalign{\smallskip}
      \hline
       \hline
      %\noalign{\smallskip}
		%	$\chi^2$ of the fit &  \multicolumn{15}{c}{$  $ } \\ 		
     %\noalign{\smallskip}
     \noalign{\smallskip}
     	\textbf{Final $k=R_p/R_s$}  & \multicolumn{6}{c}{\textbf{0.1246 }} & \multicolumn{3}{c}{ \textbf{0.1156}} & \multicolumn{3}{c}{\textbf{0.1107}} &\multicolumn{3}{c}{\textbf{0.1215}}\\
			\noalign{\smallskip}
     1 $\sigma$ (formal)  & \multicolumn{6}{c}{$\pm$0.0037} & \multicolumn{3}{c}{$\pm$0.0026} & \multicolumn{3}{c}{$\pm$0.0011} &\multicolumn{3}{c}{$\pm$0.005}\\
			\noalign{\smallskip}
			
	2 $\sigma$ (incl. syst.)   & \multicolumn{6}{c}{$\pm$0.0074} & \multicolumn{3}{c}{$\pm$0.0052} & \multicolumn{3}{c}{$\pm$0.0022} &\multicolumn{3}{c}{$\pm$0.01}\\
			\noalign{\smallskip}

	1 $\sigma$ (theo. noise limit)   & \multicolumn{6}{c}{$\pm$0.0017} & \multicolumn{3}{c}{$\pm$ 0.0012} & \multicolumn{3}{c}{$\pm$0.0007} &\multicolumn{3}{c}{$\pm$0.002}\\
			\noalign{\smallskip}
%   0.00541500   0.00259400   0.00109400   0.00512400
%     0.0108300   0.00518800   0.00218800    0.0102480
      \hline
  %    \hline	
			\end{tabular}
      \begin{list}{}{}
      \item \textbf{Notes.} {$^*$EXONEST did not fit the epoch explicitly but instead the mean anomaly at epoch}
      %\item \textbf{Notes.} Period P taken from www.exoplanet.eu; cf. \cite{bean11} therein: $P = 1.58040481 \pm 1.2\times\,10^{-7}                     \,\mathrm{days}$. $a/R_s$, $i$ taken from \cite{bean11}; b is calculated from inclination i and $a/R_s$ with a circular orbit assumption;				\textbf{\cite{bean11} refer to \cite{bean10} but I still cant find the errors in that paper. I wrote Jacob about this. Interestingly they also refer to \cite{berta11} in that paper which has $a/R_s = 14.93 \pm 0.24$ and $i=88.80_{-0.2}^{+0.25}$.}}
      \end{list}
\end{table*}

%                                					 Two column figure 
%----------------------------------------------------------- 
\begin{figure*}[ht] 
\centering
\includegraphics[width=0.75\textwidth]{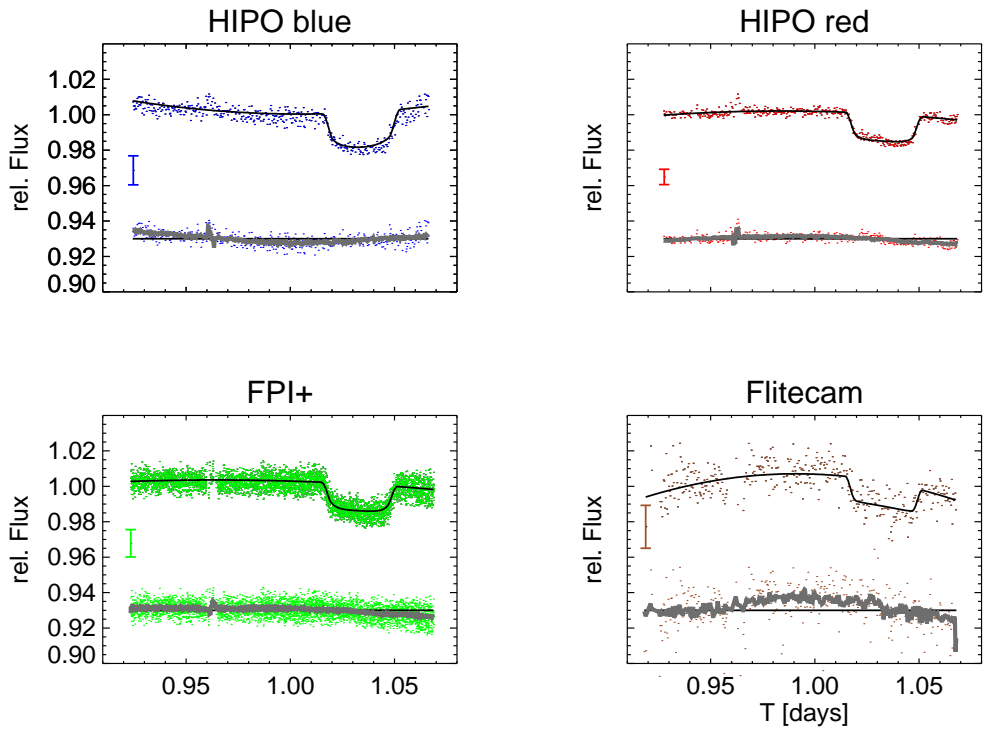} 
\caption{Initial fits (transit model + quadratic) to the raw data (outliers removed) in black. The later applied principal component noise model (grey) is overplotted to the residuals. Top left, blue: HIPO blue; top right, red: HIPO red; bottom left, green: FPI+; bottom right, brown: FLITECAM.}\label{fig:lc_raw}
\end{figure*}

%                                					 Two column figure 
%----------------------------------------------------------- 

\subsubsection{EXONEST combined with principal Component Analysis}\label{sec:PCA}

We also tested an alternative fitting and decorrelation methodology analogous to the one used in the first SOFIA exoplanet observation \cite{angerhausen15}, where they used TAP \citep{Gazak12} build on EXOFAST \citep{eastman13} in two steps combined with an intermediate decorrelation of the (airborne) observational parameters (see e.g. Figure \ref{fig:hk_plot}). Here we use the same approach, but replaced  TAP by a Bayesian nested sampling fit with EXONEST.

EXONEST is a Bayesian inference tool aimed at characterising exoplanets through Bayesian model selection, and parameter estimation (\cite{placek14, placek+14, placekknuth15,placek+15}).This tool allows one to analyse an assortment of exoplanetary data using a variety of inference engines such as Nested Sampling \citep{SiviaSkilling06}, MultiNested Sampling \citep{Feroz+09, Feroz+11, Feroz+13}, Metropolis-Hastings Markov chain Monte Carlo (MCMC) Sampling \citep{Metropolis+53}, and Simulated Annealing \citep{otten89}. MultiNest was chosen in this specific analysis for its efficiency in sampling from complicated parameter spaces. Inputs to EXONEST consist of the prior probabilities for each model parameter, which reflect ones knowledge about the model parameters prior to having analysed the data, and the likelihood function, which depends on the model and the expected nature of the noise.

The four channels of photometric time series obtained from SOFIA were simultaneously fit using the model of \citet{mandel02}, which is parametrised by the planet-to-star radius ratio, and quadratic limb-darkening coefficients for each channel, the scaled semi-major axis, $a/R_\star$, and the impact parameter $b = \frac{a}{R_\star}\cos i$. The planet-to-star radius ratio, quadratic limb-darkening coefficients, and the impact parameter were each sampled from uniform prior probability distributions over the ranges [0, 0.2], [0,1], and [0,1], respectively. For more straightforward comparisons to other methods, we again fixed the scaled semi-major axis at $a/R_s = 14.97$, the impact parameter at $b = 0.277026$, the orbital period to $P = 1.5804055929$ days, and the eccentricity to zero. Assuming the noise in each channel to be Gaussian distributed, the likelihood function for each channel, $L$, takes on the form:

 \begin{equation}
	L = \sum_{i=1}^N \frac{1}{\sqrt{2\sigma_i^2}} exp\left(  -\frac{(F_i - d_i)^2 }{2 \sigma_i^2} \right) 
 \end{equation}

where N is the number of data points in the channel, $\sigma_i$ is the standard deviation of the $i^{th}$ data point $d_i$, and $F_i$ is the corresponding model prediction.  
MultiNest works to maximise the likelihood (or log-likelihood) function to ultimately obtain the posterior distribution from which parameter estimates can be derived. 

%\subsubsection{principal Component Analysis}
Prior to fitting, a 3-$\sigma$ clipping was performed on the raw data for outliers removal. Following the method in \citet{angerhausen15} the transits were then modeled in three steps. First, the raw light curves were fit with a transit model and a 2nd-order polynomial to account for airmass. The residuals to the initial fits were then modeled with the first 16 principal components $p_i(t)$ in order to decorrelate with the observational parameters, sampled at the same time as our exposures, as a linear combination $R_{model}(t)=\sum c_i \times p_i(t)$. Figure \ref{fig:lc_raw} displays the raw data with the transit+quadratic fits, and the corresponding residuals (again including the quadratic) with these principal component fits. 
This noise model was computed independently from the iterative EXONEST analysis and no marginalisation has been done over the instrument model correlation terms. As also argued in \cite{angerhausen15} we choose this approach because the lack of post transit baseline causes convergence issues and a high risk of running into degeneracies between the noise model and the actual transit depth.

Finally, the best-fit principal component model for each channel was subtracted from the raw data, and the (decorrelated) transits were fit again. 
The results of these simulations are displayed in Table \ref{tab:lc_fits}, and the corresponding principal component analysis (PCA) noise models are shown in grey in Figure \ref{fig:lc_raw}.  
The estimated planet-to-star radius ratios are in good agreement with the results from TLCM in Section \ref{sec:TLCM_modelling} except for the FLITECAM observations, which differ by slightly more than two-sigma. We discuss this discrepancy in Section \ref{sec:noise_ana} and take this as an argument to report the 2 $\sigma$ error as a reflection of the real systematic noise in addition to the formal 1 $\sigma$ error.

\begin{figure*}[!ht]
 \centering
  \includegraphics[width = 0.85\textwidth]{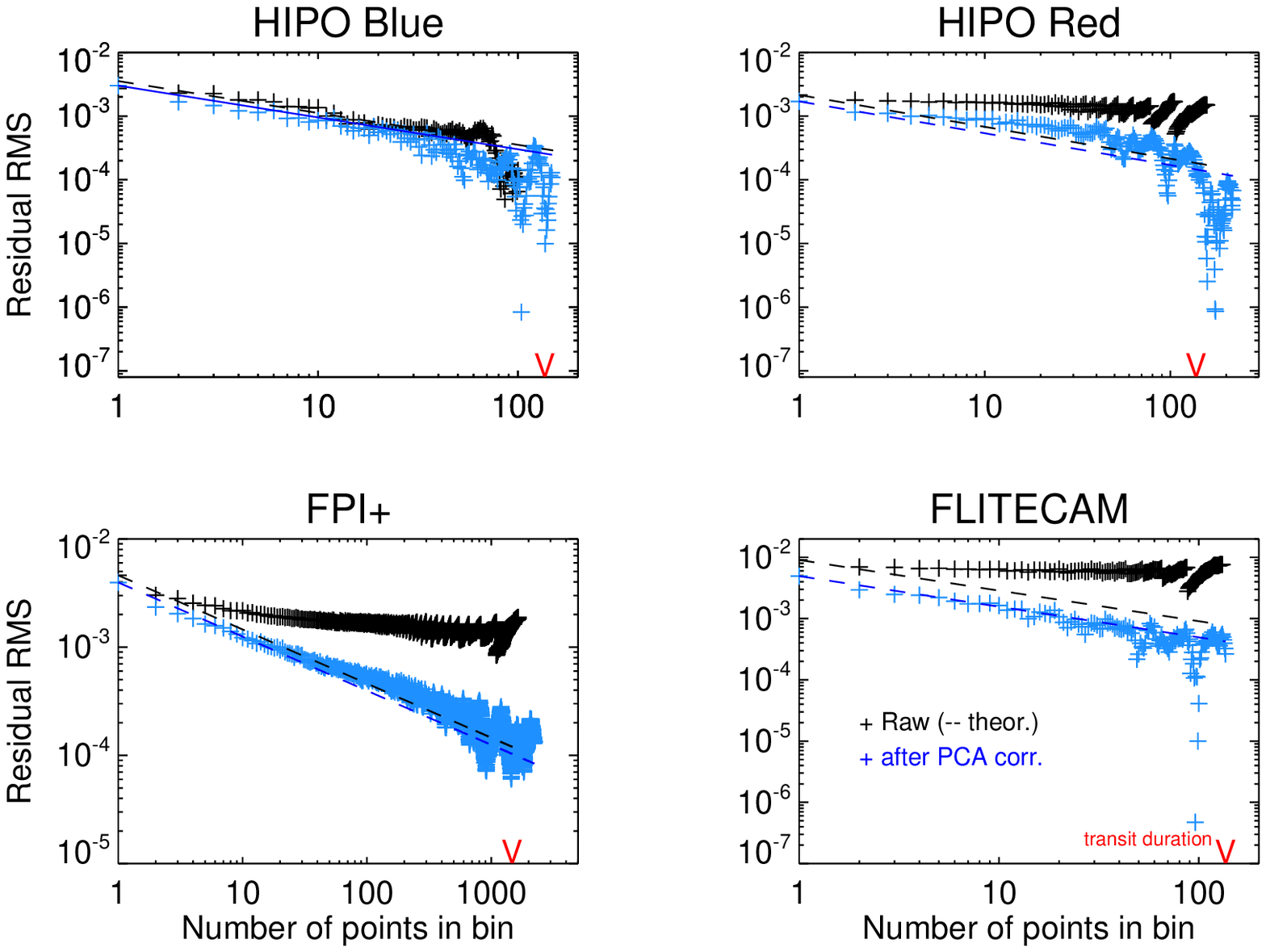}
  \caption{Plots of the residual rms versus bin size for each channel. The black marks represent the values for the raw light curves, whereas the blue corresponds to the de-trended light curves after the best-fit principal component model was subtracted. Dashed lines represent the theoretical values for pure white noise, red arrows mark the transit duration as reference.}\label{fig:Allan_PCA}
\end{figure*}

\subsubsection{Correction of twilight effects on HIPO-blue light curve}\label{sec:HipoCorr}
\begin{figure*}[!ht] 
\centering
\includegraphics[width =0.48\textwidth]{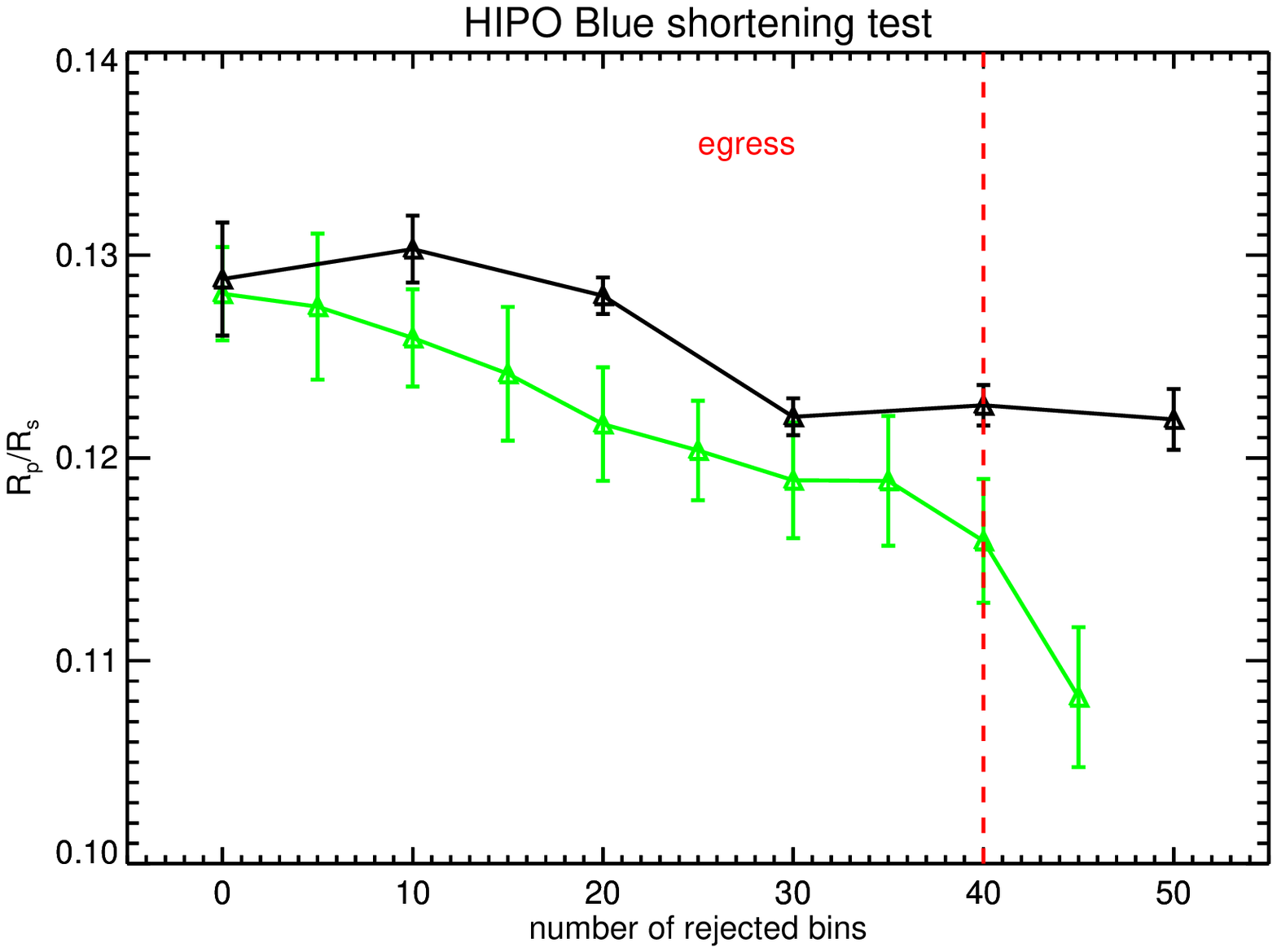}
\includegraphics[width =0.48\textwidth]{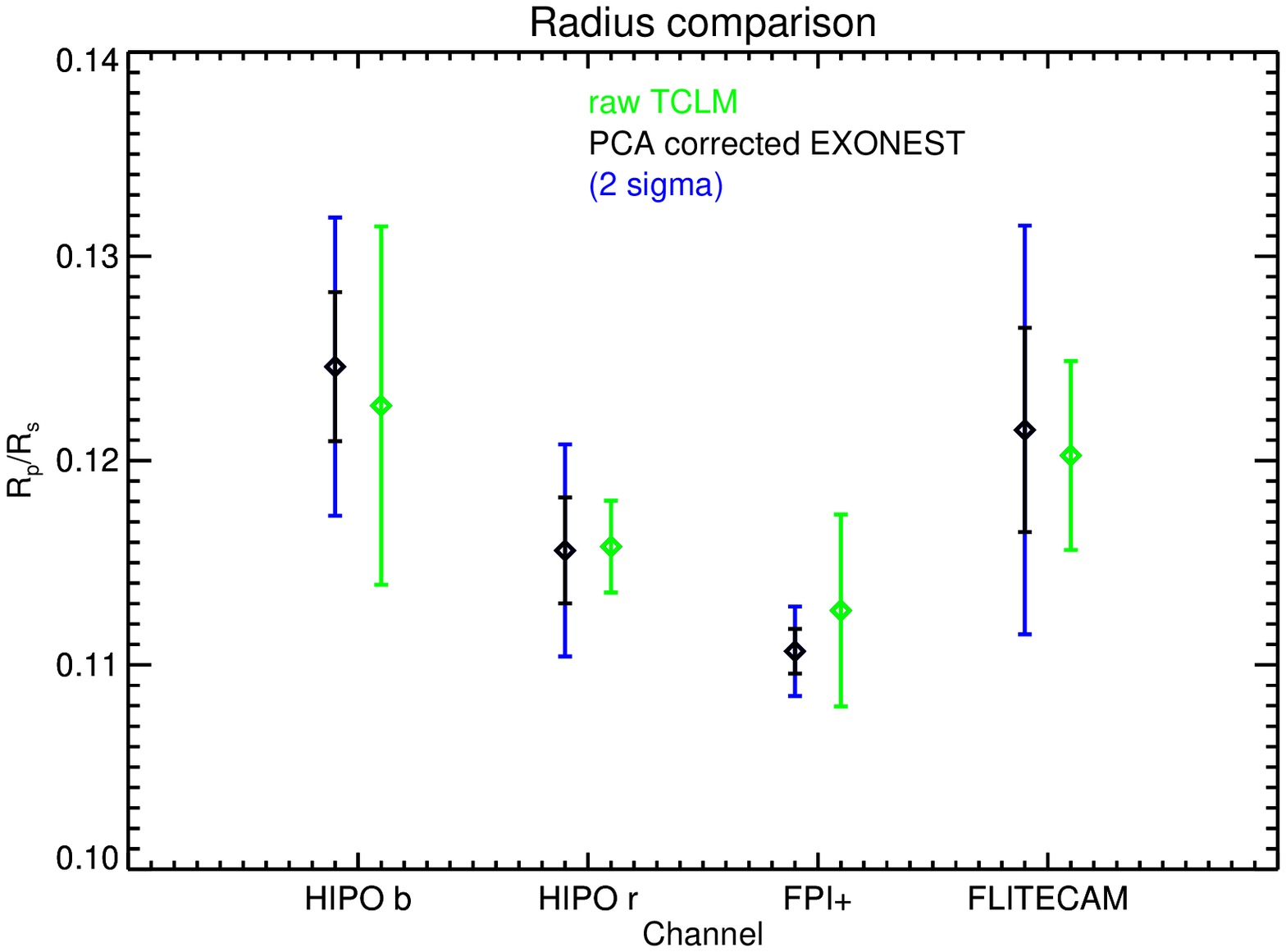}
\caption{Left: Estimates of the planet-to-star radius ratio for the the HIPO blue observations; from left to right, more data points were neglected from the post-transit observations in steps of 5/10 points. The vertical red line indicates the end of egress, i.e. neglecting more points would neglect in transit data.  This indicates that the upward post-transit trend seen in Figure \ref{fig:lc_raw} and has affected the measured transit depth.  Right: Comparison of the results obtained with our two different fitting approaches.}
\label{fig:HIPO_chopped} 
\end{figure*}

In both reductions, the methods used to account for systematic noise (polynomial slope, linear combination of principal components) failed to account for an upward trend in the HIPO-blue post-transit observations. These observations were taken close to sunrise and affected the shortest wavelength channel significantly towards the end of the flight. This led to a transit depth several standard deviations from previous measurements. To investigate whether this post-transit slope was skewing the estimated transit depth, a series of simulations were performed with an increasing amount of post-transit data neglected. The results of this process are displayed in the left-hand panel of Figure \ref{fig:HIPO_chopped}. For the PCA corrected data, after thirty points were neglected from the post-transit observations, the derived value for the transit depth began to plateau. The resulting value associated with thirty post-transit points neglected is $R_p/R_s = 0.122537 $ for the PCA corrected data. For the TLCM fits to the raw data the results did not plateau but instead consistently decreased as more data points were neglected to a final as low as $R_p/R_s$ =0.11845. In a conservative approach we therefore decided to report the whole range of possible outcomes as our final values for HIPO blue in both methods (see left panel of Figure \ref{fig:HIPO_chopped} and Table \ref{tab:lc_fits}) and again with the 2 $\sigma$ error to account for the systematic contribution in addition to the formal 1 $\sigma$ error.

\subsection{Influence of host star activity} \label{sec:activity}
Star spots not crossed by the transiting planet cause the average brightness to be higher along the transit chord than on the rest of the stellar hemisphere, which leads to an overestimation of the planet-star radius ratio \citep{csizmadia13b, czesla09}. This effect is wavelength dependent due to the different temperatures and spectral energy distributions of spot and unspotted photosphere \citep{pont08,sing11}. The host star GJ1214 was photometrically monitored in the observing season 2014 with the robotic telescope STELLA and its imager WiFSIP \citep{strassmeier04}. The observations continued the WiFSIP monitoring program of 2012 and 2013 presented in \citet{nascimbeni15}. Details of the observations and data reduction are presented in Mallonn et al. 2017, in prep. The long-term photometry proved the host star to be at maximum brightness at the time of the SOFIA transit observation. If maximum brightness was interpreted as a spot-free visible hemisphere, no correction of the derived transit parameters would be needed. However, the monitoring only yields information on the relative change in spot filling factor, but no information about the level of spots permanently visible. \cite{nascimbeni15} estimated the differential correction for a filling factor of 2\% between Bessel B and Bessel R to $<0.0001$ in the planet-star radius ratio k. If we conservatively assume a permanent spot filling factor of 4\% (which is on the order of the maximum change in the spot filling factor in the season 2014), it results in a differential spot correction of $~0.0002$ in the optical, which is an order of magnitude smaller than our error bars for k. Since the value of the correction further decreases towards the NIR, we conclude that a correction for un-occulted spots is negligible in our case. We note that the monitoring light curve of 2014 displays the largest amplitude and longest apparent periodicity ever observed for the super-Earth host star GJ1214. We refer to Mallonn et al. 2017, in prep. for an in-depth analysis of five years of GJ1214 monitoring from 2012 to 2016.

\subsection{Noise analysis}\label{sec:noise_ana}
%\subsubsection{1/f noise and correction of correlated parameters}\label{sec:noise_1f}
In Figure \ref{fig:Allan_PCA} we show the variance of the residuals as a function of bin size. This scheme is
commonly used in the literature to assess the amount of correlated noise and as a visual test of the noise correction method. While the raw data in both plots show strong deviation from the theoretical limit (dashed lines, $\sigma^2 (l) = \sigma^2(0)/l$) for pure white noise, they show that the PCA corrected data is much closer to the expected line. The red lines in these figures mark the length of the transit and are the frequencies that eventually bias the results most.
 This improvement shows that a large amount of the time-correlated components in the time series of residuals were removed after the correction via PCA and could explain the discrepancy with the TLCM fitted results. However, \citet{Cubillos17} report a number of caveats for using variance plots as measure for residual correlated noise, as well as for other frameworks dealing with systematic noise such as the residual-permutation or wavelet-likehood methods.
 
In an alternative hands-on test to explore how close we approach the photon noise limit we simulated a data set with white noise added according to the values derived from the raw data.
As shown in Table \ref{tab:lc_fits} we obtain sensitivities of 500 ppm/min for FPI+, 
400 ppm/min for HIPO-Red, 1000 ppm/min for HIPO blue and 1000 ppm/min for FLITECAM    
using the gain values provided by the instrument teams. After running the same reduction and fitting procedure with the theoretical data, we obtain a factor of 2.2/2.2 for HIPO (blue/red) and 1.6 for FPI and 2.5 for FLITECAM smaller error bar for the derived transit depth compared to the formal 1-$\sigma$ error of the real data fits for the transit depth. This is consistent with the findings in \cite{angerhausen15}, where they reached $\sim 2$ times the photon noise. We decided to additionally report the 2 $\sigma$ error with our final result to account for any residual systematic.

\section{Transmission Spectra Modelling}\label{sec:atmos_modelling}
\begin{figure*}[!ht]
     \centering
     \includegraphics[width=0.9\textwidth]{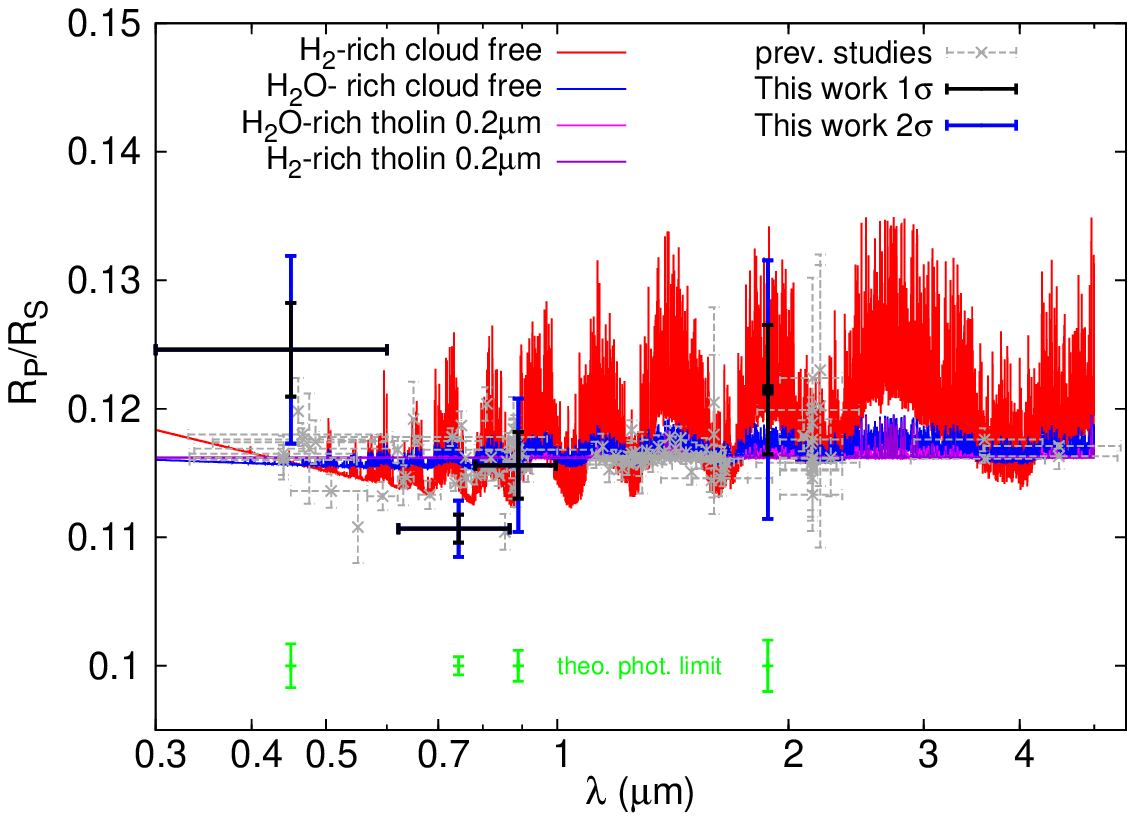}\\     
     \caption{\textbf{Model transmission spectra of GJ 1214b, previous measurements (grey) and this work (black, blue).}
		 \textit{Observations:} \citet{bean10,bean11}; \citet{croll11}; \citet{crossfield11}; \citet{desert11}; \citet{berta12}; \citet{demooij12}; \citet{colon13}; \citet{demooij13}; \citet{fraine13}; \citet{caceres14}; \citet{kreidberg14}; \citet{wilson14}; \citet{nascimbeni15} and \citet{Rackham2017}.}
     \label{fig:atmos_spectrum}
\end{figure*}

% \subsection{Modelling Studies Related to GJ1214b}
%Introduction\\
%Model studies of these planets aim to interpret the observational data (e.g. T, (bulk-)composition, albedo, hazes etc.) which is starting %to be gathered.\\
%Since the water vapor absorption bands in the near infrared are completely dampened by the clouds, it seems unlikely that one may derive the mean molecular mass of the atmosphere from any absorption features. However, measuring the slope of the atmospheric absorption at shorter wavelength may allow to distinguish between different Rayleigh scattering efficiencies due to different mean molar masses of the atmospheres, hence water or hydrogen dominated atmospheres, or even Mie scattering by larger particles (see e.g.~\cite{howe12} and \cite{benneke13}). \\
For the calculation of the theoretical transmission spectra we follow the methods described in \cite{Gaidos2017}. We adopt selected scenarios from \citet{kreidberg14}, namely a water-rich case with 99\% \ce{H2O} and 1\% \ce{H2} \& \ce{He}, as well as a hydrogen-dominated atmosphere composed of 99\% \ce{H2} \& \ce{He} and 1\% water.
The water opacity is calculated with the HELIOS-k \citep{grimm15}, employing the HITEMP2010 line list \citep{Rothman2010}. Collision induced absorption from HITRAN2010 is used for \ce{H2}-\ce{H2} and \ce{H2}-\ce{He} collisions. 
The molecular scattering cross sections are derived via the Rayleigh scattering equation
\begin{equation}
  \sigma_\mathrm{rayleigh} = \frac{24 \pi^3 \nu^4}{n_\mathrm{ref}^2} \times \left(\frac{n(\nu)^2 - 1}{n(\nu)^2 + 2}\right)^2 \times K(\nu) \ ,
\end{equation}
where $\nu$ is the wavenumber, $n$ the refractive index, $n_\mathrm{ref}$ a reference particle number density, and $K$ the King factor.
The corresponding data for \ce{H2} is taken from \citet{cox00}, for \ce{H2O} from \citet{wagner2008international} and \citet{Murphy1977}, and from \citet{sneep05} in case of \ce{He}. For the scenarios that include high-altitude hazes, we assume that the haze particles are composed of small hydrocarbon clusters \citep{kreidberg14}. The optical constants of these tholins are taken from \citet{Khare1984}. The cloud pressure is 0.001 mbar for the hydrogen-dominated case and 0.002 mbar for the water-dominated scenario, respectively.
Figure \ref{fig:atmos_spectrum} shows the resulting transmission spectra for the described scenarios with data from the literature and the radius ratios at different wavelengths obtained by the SOFIA measurements. \\
The results clearly suggest that the spread of the obtained photometric data points by SOFIA is larger than the spread in the theoretical transmission spectra. Especially, the very large planetary radius obtained with HIPO blue seems to suggest the presence of a Rayleigh slope which contradict the apparent flatness of the planet's spectrum reported by other studies \citep[e.g.][]{kreidberg14}. While within the 2$\sigma$ error bars, most photometric points correspond roughly to the simulated spectra. The photometric precision of our SOFIA measurements are unfortunately not good enough to put better constraints on the atmospheric composition than previous studies with similar sensitivities.

%Modelling Temperature - most studies so far adopt a 1D simplified form of the radiative transfer equation (see e.g. \cite{guillot10, miller-ricci09}) 
%which assume hydrostatic equilibrium, a constant opacity in the visible and infra-red and an internal energy source imposed at the model 
%lower boundary. Results for GJ1214b suggest a mostly isoprofile temperature in the range (500-600K) for pressures smaller than $\approx 1\,\mathrm{bar}$.\\
%\\
%Modelling Composition\\
%It is convenient to consider three atmospheric regions - first, an upper region (for $p~<0.1\,\mathrm{bar}$) where composition is controlled
%by photochemistry; second, a middle region (which extends down to a few bar) where composition is controlled by mixing and thermal quenching and
%third, a lower region where composition is characterised by thermodynamical equilibrium. Model studies aim simulate the composition of all three regions
%in a coupled framework. Important unknowns are e.g. the trace constituents, the Eddy mixing coefficients and the assumed $C/O$ ratio.
%\cite{fortney13} discussed the possible hazes on GJ1214b via formation of higher hydrocarbons.
%\cite{miller-ricci12} discussed the effect of e.g. atmospheric vertical mixing on composition, hence upon GJ1214b's transmission spectrum.
%\cite{hu14} discussed possible hydrocarbon features in the transmission spectrum of GJ1214b.

%                               			 two column figure (place early!)
%-----------------

%
%
%______________________________________________________________
\section{Conclusions}\label{sec:conclusion}
  
\subsection{Summary and Results}
We used FLIPO and FPI+ on board of SOFIA to simultaneously observe a transit of the Super-Earth GJ 1214b in three optical ($open blue= 0.3-0.6\,\mathrm{\mu m}$, $i'= 0.8\,\mathrm{\mu m}$, and $z'=0.9\,\mathrm{\mu m}$) and one never before covered infrared channel (Paschen-$\alpha$ cont. $1.9\,\mathrm{\mu m}$) and present the light curves and corresponding transit depths in these bands. 
We compare to previous observations and state-of-the-art synthetic models of its atmosphere. Unfortunately our results are not sensitive enough to constrain the models any better than previous observations already did. As discussed in Section \ref{sec:datareduction} and \ref{sec:modelling} some of our channels are dominated by residual systematic noise. However, the results can become useful in combination with prior and future observations for future retrievals.
We found that the PCA is a powerful tool to reduce the correlated noise in SOFIA data and recommend its usage for future data analysis. In our present case of GJ 1214b PCA was able to reduce the noise level by two orders of magnitude and finally we reached a noise level of two times of the photon noise. The data also presents a second reference for exoplanet transmission spectrophotometry with SOFIA and the first in all four available channels. In Fig. \ref{fig:atmos_spectrum} we show the theoretical limits for our observation (green bars). With further improvements to our calibration strategy and a better understanding of our instruments we are confident that we can get closer to these limits and make SOFIA more competitive in this field. We summarise SOFIA prospects in the last section.

\subsection{Exoplanets with SOFIA}
In this paper we presented the first exoplanet transit observation with SOFIA that leveraged all four possible  channels for simultaneous spectrophotometry. While two of the optical instruments produced good results, the IR channel did not reach the expected sensitivity. Our measurements suffered from insufficient calibration files for the FLITECAM channel. The challenge is that SOFIA does not have a sufficiently bright flat field source for use with such a narrow band filter as the 1.9 $\mu m$ Paschen-$\alpha$ continuum filter. Another lesson learned from this flight is to avoid only short baselines before or after transit at all costs. As the problems with the HIPO blue channel here shows, it is crucial to have at least 30-60 minute baseline before and after the occultation to be able to trace systematic changes and correct for instrumental or other observational effects. For future observations we therefore recommend a more careful calibration scheme in particular for obtaining flat fields and if possible a flight plan that allows for more time before and after the transit.

However, even in the current configuration and with all these constrains there are certain niches that we were able to identify with this and the previous \citet{angerhausen15} SOFIA exoplanet observation.

In summary this phase space is:
\begin{itemize}
    \item  bright host stars (like HD 189733b) - for which \citet{angerhausen15} demonstrated the ability to perform absolute optical photometry  
 	\item  short transit durations, 
 	\item  science cases that leverage SOFIA's unique  capability to observe IR/OPTICAL simultaneously - which complements James Webb Space Telescope (JWST) coverage and/or can be used for JWST target selection  and support
 	\item and transits that are rare/time-critical and require a deployment 
\end{itemize}

With the upcoming TESS (Transiting Exoplanet Survey Satellite) \citep{sullivan15} and PLATO (PLAnetary Transits and Oscillations of stars) \citep{rauer14} missions we will see a lot more transiting exoplanets that fall into these categories. Furthermore it is possible to update SOFIA's instrumentation with a modernised VIS/NIR precision photometer similar to the previously proposed NIMBUS concept \citep{2012SPIE.8446E..7BM}. This accompanied by a reliable and robust water vapor monitoring system, could make important SOFIA-unique contributions to exoplanet science.

\begin{acknowledgements}
The authors would like to thank the anonymous referee for her/his valuable comments  which substantially helped improving the quality of the paper. 
This research has made use of SIMBAD, 2MASS, GCVS catalogue, and AAVSO variable search index. 
D.A. acknowledges the USRA NASA postdoctoral program and  the support of the Center for Space and Habitability of the University of Bern. S.C. acknowledges the Hungarian OTKA Grant K113117. 
D.K. gratefully acknowledges the support of the Center for Space and Habitability of the University of Bern and the MERAC Foundation for partial financial assistance. This work has been carried out within the frame of the National Centre for Competence in Research PlanetS supported by the Swiss National Science Foundation. D.K. acknowledges the financial support of the SNSF.
HIPO work at Lowell is supported by USRA subcontract 8500-98-003. 
FLITECAM work at UCLA is supported by USRA subcontract 08500-05, PI Ian McLean. 
M.G. acknowledges financial support by the Helmholtz association, PD-015, and the Technische Universit\"at Berlin.
\end{acknowledgements}

\bibliographystyle{aa} % style aa.bst
\bibliography{literature_sofia} % your references Yourfile.bib

\end{document}